# Bi-Critical Central Point Of $J_{FN}$-$J_{SN}$ Ising Model Phase Diagram

**Y. Boughaleb[1,2], M. Noureddine[1] and R. Nassif [2,1]**

[1]Université Hassan II, Faculté des Sciences Ben M'sik, Casablanca Maroc

[2]Université Chouaib Doukkali, Faculté des Sciences El Jadida Maroc

**Abstract:**

When interfaces between ordered domains are ordered clusters, frustration disappears. A phase with mixed ordered structures emerges but no length scale can be associated to. We show that sum of densities of each structure plays the role of order parameter. For, we consider a regular half full lattice with repulsive interaction extended to second neighboring particles. Ordered structures are *p(2X2)* when ratio between second and first neighboring interaction energies $R=J_{SN}/J_{FN}<0.5$ and degenerate *p(2X1)/p(1X2)* for R>0.5. The ground states coexist with another named *p(4X2)/p(2X4)* at central bi-critical point: a state to cross when passing between non-frustrated ordered phases.

The study of frustrated ferromagnetic systems always arouses the interest [1-6]. Magnetic systems elaborated show that the nature of the ordered magnetic phases at low temperatures depends on the concentration of the magnetic species and on the relative magnitude of competing energy scales [7-8]. This confirms that these compounds are well described by theoretical discrete models considering competition between the interaction energies of the particles nearest neighbors. The recent discovery of high-Tc superconductivity in pnictides [9-10] shows that $J_{FN}$-$J_{SN}$ Ising model offers a valuable framework for dealing with the phenomena involved. However, the approximations adopted to solve the model gives rise to a controversial phase diagram [11-15]. In fact, it was shown that square lattice presents superantiferromagnets $p(2X2)_+$/ $p(2X2)_-$ at the coverages $\theta=1/4$ and $\theta=3/4$ respectively [11;16]. When half occupied, there appears in the low temperature limit a *p(2X2)* ordering phase or a degenerate *p(2X1)/p(1X2)* ordered structure for $R<0.5$ and $R>0.5$ respectively. In both cases, characterization of phase transition is done by means of order parameters that quantify the breakdown of lattice occupation symmetry. Nowadays everyone agrees that the transition line between disorder and ordered phase for R <0.5 is continuous but the situation $R>0.5$ is still under debate; particularly the extent of interaction rate for which transition is first order [1;13-15;17]. The system with $R=0.5$ is equally confused. Here the two phases do have the same energy as well as a third "2d-honeycoumb *p(4X2)/p(2X4)*" structure joining a central particle to one first and to two second neighboring particles. In the absence of third neighboring interaction energy Monte Carlo computations using metropolis algorithm cannot allow complete relaxation of the system as jump of unique particle is far from altering the stability of clusters formed. The only one way to bypass the status quo observed is to improve the relaxation rate. This was done by considering parallel tempering method [18]. Calculations determine critical temperatures for lattices with limited sizes but finite size scaling conclude an absence of transition in the thermodynamic limit [14]. Dealing with such situation, we consider a bidimensional layer occupied by repulsively interacting particles. we consider that no pure phase holds but large ordered clusters. Here frustration which usually comes from antiphase boundaries disappears. In fact, interfaces between two ordered domains are simply clusters of the third ordered structure. For we choose an order parameter which quantifies densities of existing structures in the lattice and whom fluctuations determine critical temperature. The grandeur is then generalization to all R value. Phase diagram is deduced and criticality discussed.

The study is performed in the framework of lattice gas model. We consider a bi-dimensional square lattice with lateral size $L$ for which interaction energy is insured by the following Hamiltonian of static interactions:

$$H = -\frac{1}{2} J_{FN} \sum_{\langle i,j \rangle} n_i n_j - \frac{1}{2} J_{SN} \sum_{\langle\langle i,k \rangle\rangle} n_i n_k - \mu \sum_i n_i \qquad (1)$$

"$n_i$" is a Boolean occupation number; $R$ denotes the ratio between interaction energies of second $J_{SN}$ and first $J_{FN}$ neighboring particles, $\mu$ is the chemical potential and number of brackets indexes the neighbors.

Monte Carlo computations allow relaxation towards equilibrium insured meaning standard Metropolis transition rate applied to jump of single particle to its immediate surroundings. The lattice is periodically bounded and computation of physical grandeur is performed after complete relaxation.

As mentioned the low temperature regime is characterized by an algebraic order parameter "*AOP*" that measures densities of ordered structures in the lattice $\rho = \sum_i \rho_i$ . "$i$" designates the structure and $\rho_i = \sum n_{\vec{r}} / L^2$ where summation is performed once constraints relative to establishment of a type of order are fulfilled. The ability of describing the order/disorder transition by the quantity this defined is tested by Grand Canonical Monte Carlo simulations. Fig. **1** reports adsorption isotherm as well as the behavior of *AOP* for different temperature regimes. For high temperatures adsorption isotherm presents Langmuir gas behavior. Absence of ordering is characterized by weak values of *AOP*. This can be understood by existence of ordered clusters of limited sizes as a consequence of the equal probability of occupying lattice sites. To transition to ordering marked by plateau in adsorption isotherm corresponds a clear increase of *AOP* values. Signature of emergency of a kind of long range order in the low temperature regime is attributed to collapse of both adsorption isotherm and *AOP* curves in the interval for which the coverage is fixed to $\theta = 0.5$.

As done in standard studies of phase transition we compute the susceptibility of the system. The latter measures this time the autocorrelation of the introduced *AOP*:

$$\chi_T \equiv \frac{1}{K_B T}\left[\langle \rho^2 \rangle - \langle \rho \rangle^2\right] \tag{2}$$

The results confirm the transition to a low temperature phase constituted by mixture of different types of ordered clusters (see fig.2). Divergence of susceptibility gives effective critical temperature of the considered lattice size. Scaling analysis applied to the results indicates the transition is non-universal ($\nu=0.94$ and $\gamma=1.56 \pm 0.13$). Another proof of the validity of our hypothesis lies in characterizing the phase transition to all *R* values. The phase diagram of figure **3** shows that obtained critical temperatures agree with the results of studies within Monte Carlo technique [11;14;19]. Once again finite size scaling analysis confirms an Ising-like behavior with universal critical exponents of the system with $R<0.5$ and that order/disorder phase transition for $R>0.5$ is non-universal ($\nu=0.70$ and $\gamma=1726 \pm 0.08$ for $R=0.75$ for example).

Now that transition to ordered phase at $R=0.5$ has non null temperature in the thermodynamic limit, it would be interesting to characterize transition from *p(2X2)* ordered phase to *p(2X1)/p(1X2)* one as a consequence of varying the relative magnitude of competing interaction energies. For we compute the Cowley [20] short range order parameter $\alpha = \left(\left\langle n_i n_j \right\rangle - \theta\right) / \theta(1-\theta)$ versus intensity of $R$ ratio for different temperatures of the system. The situation corresponds to crossing over the phase diagram. One distinguishes plateaus in the low temperature regime (Fig. **4**) that quantify emergency of ordered structures in the lattice with a *p(2X2)* structure when $R<0.5$ and a *p(2X1)/p(1X2)* phase for $R>0.5$. Transition to disorder is signalled by intersection between the critical curve and the line of constant temperature. Extent of the plateaus increases by cooling while phase with mixed structures constitutes remarkable point. Jump in short range order parameter reflects the transition is first order. Then the central point presents bi-critical behaviour and constitutes the only one way between the two ordered pure phases.

We can conclude from the results obtained through the definition of the *AOP* that it is possible to envisage a transition to an ordered phase with a mixture of structures whenever the overlap of existing structures do not give rise to antiphase boundaries. This transition presents a bi-critical behavior depending on whether the establishment phase is carried out by cooling at $R = 0.5$ or by varying the ratio of exchange constants at fixed temperature. The transition for $R > 0.5$ is second order with non-universal critical exponents.

**Acknowledgments:**

We are grateful to ***Pr. J. Prost*** for fruitful discussions and suggestion.

**Figure captions:**

**Fig. 1:** Variation of adsorption isotherms and density of ordered structures versus chemical potential for different interaction regimes at $R=0.5$. Size of the lattice is fixed to $L=50$.

**Fig. 2:** Validity of algebraic order parameter $\rho$ hypothesis: Behavior of the susceptibility defined as fluctuations of ordered structures density versus reduced temperature for different sizes of the lattice and finite size scaling analysis (the inset).

**Fig. 3:** Phase Diagram of the half full lattice obtained by means of the algebraic order parameter fluctuations.

**Fig. 4**: Behavior of short range order parameter while crossing the phase diagram. The curve shows bi-critical character of the remarkable phase diagram central point.

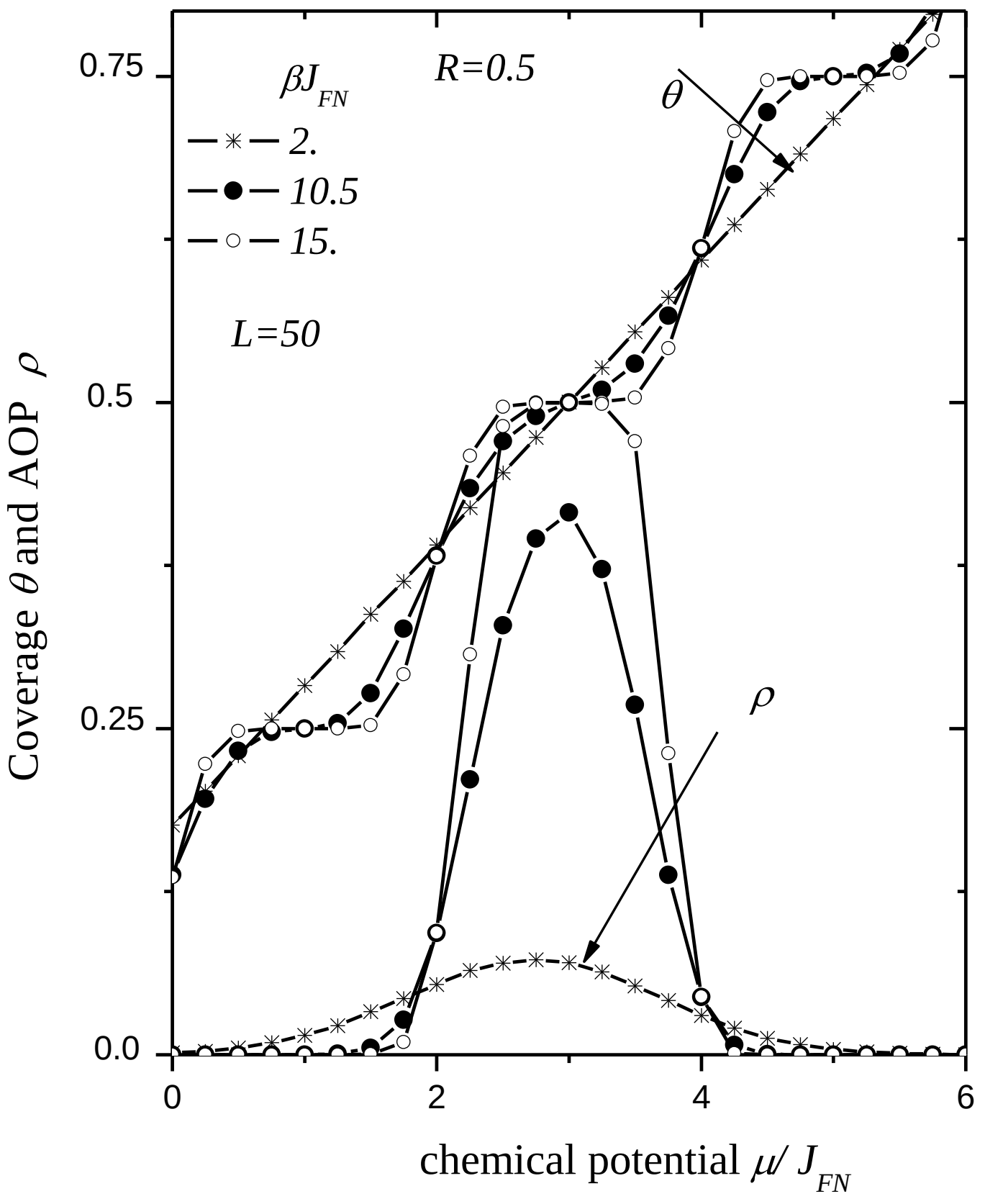

Fig. 1

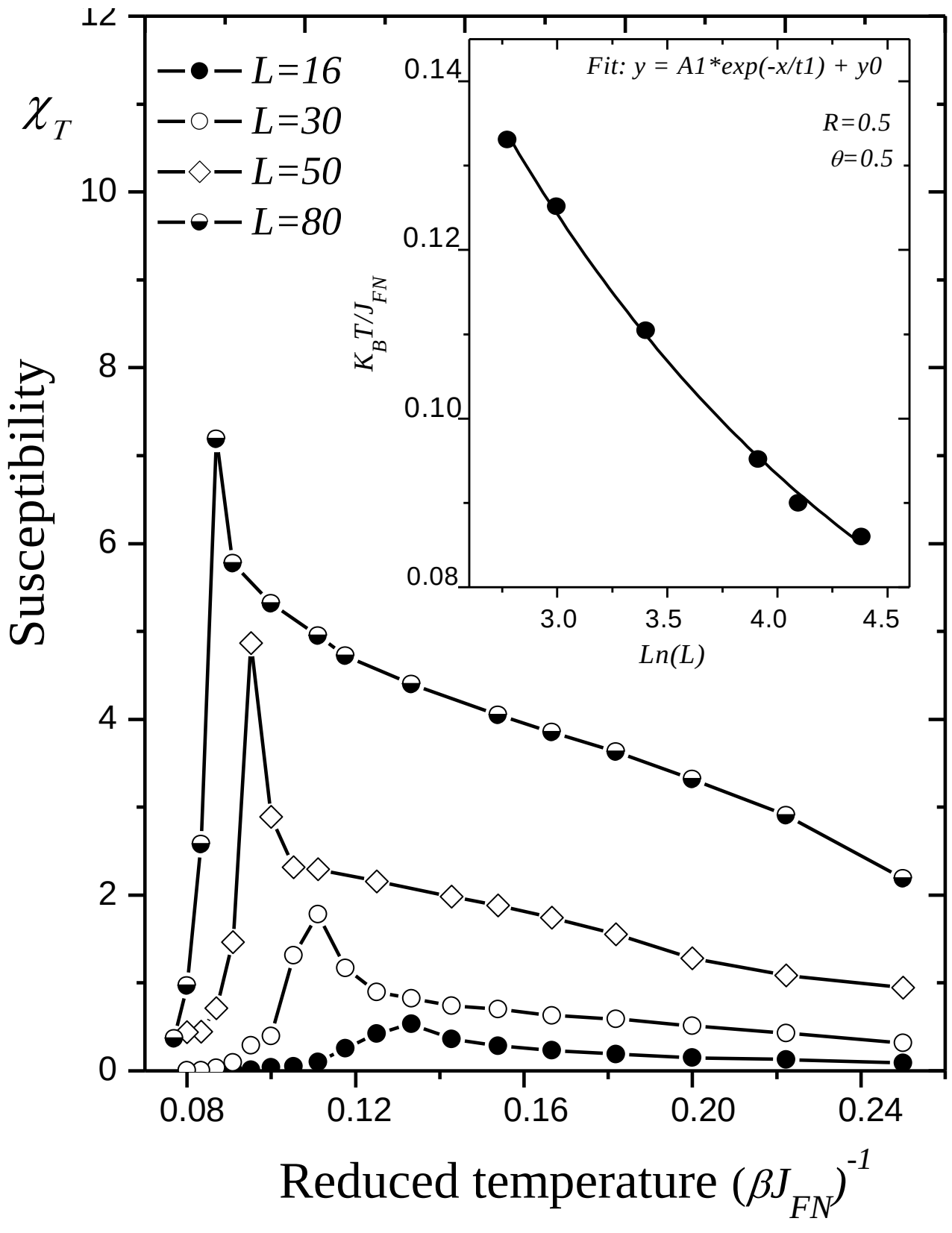

Fig. 2

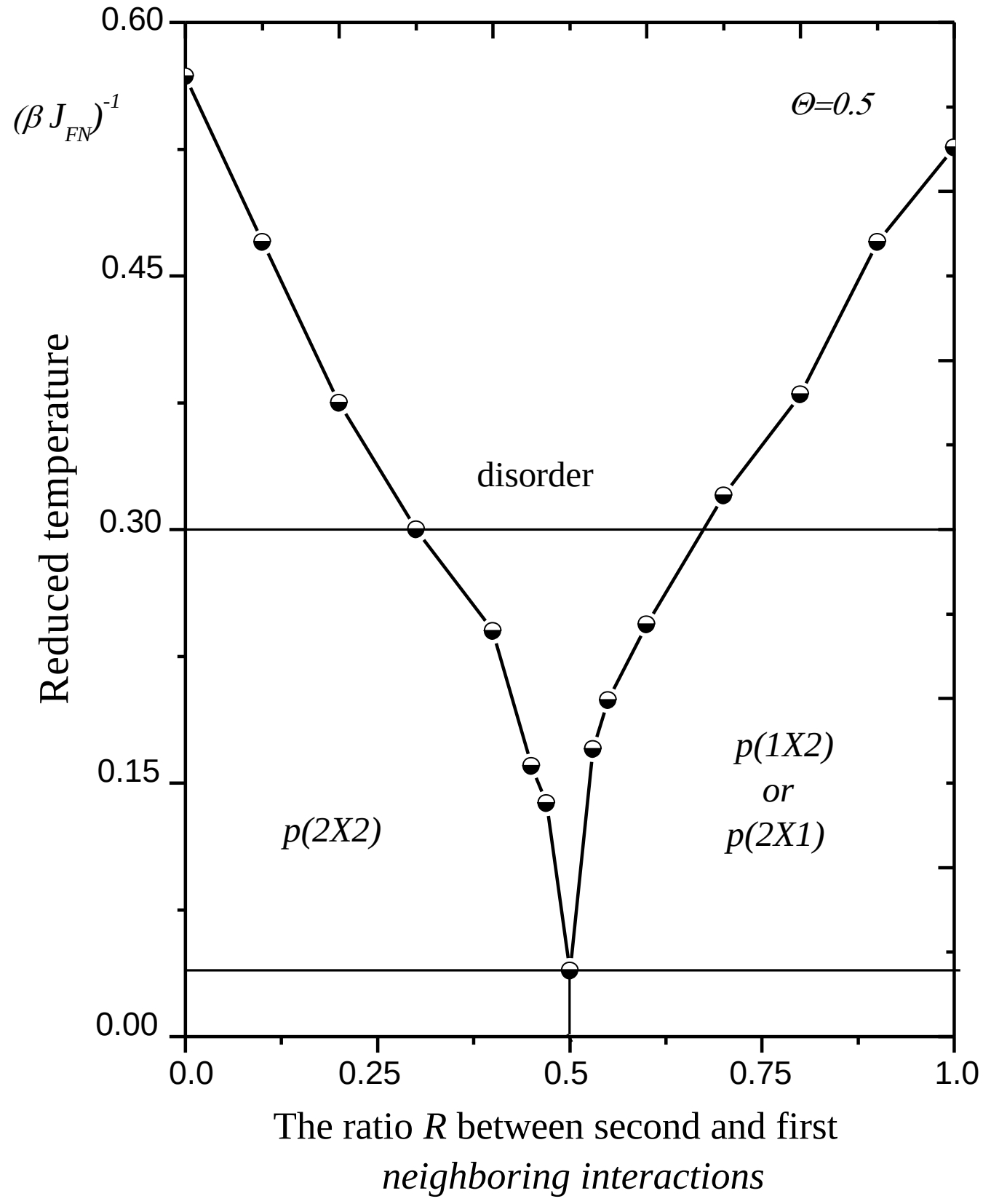

Fig. 3

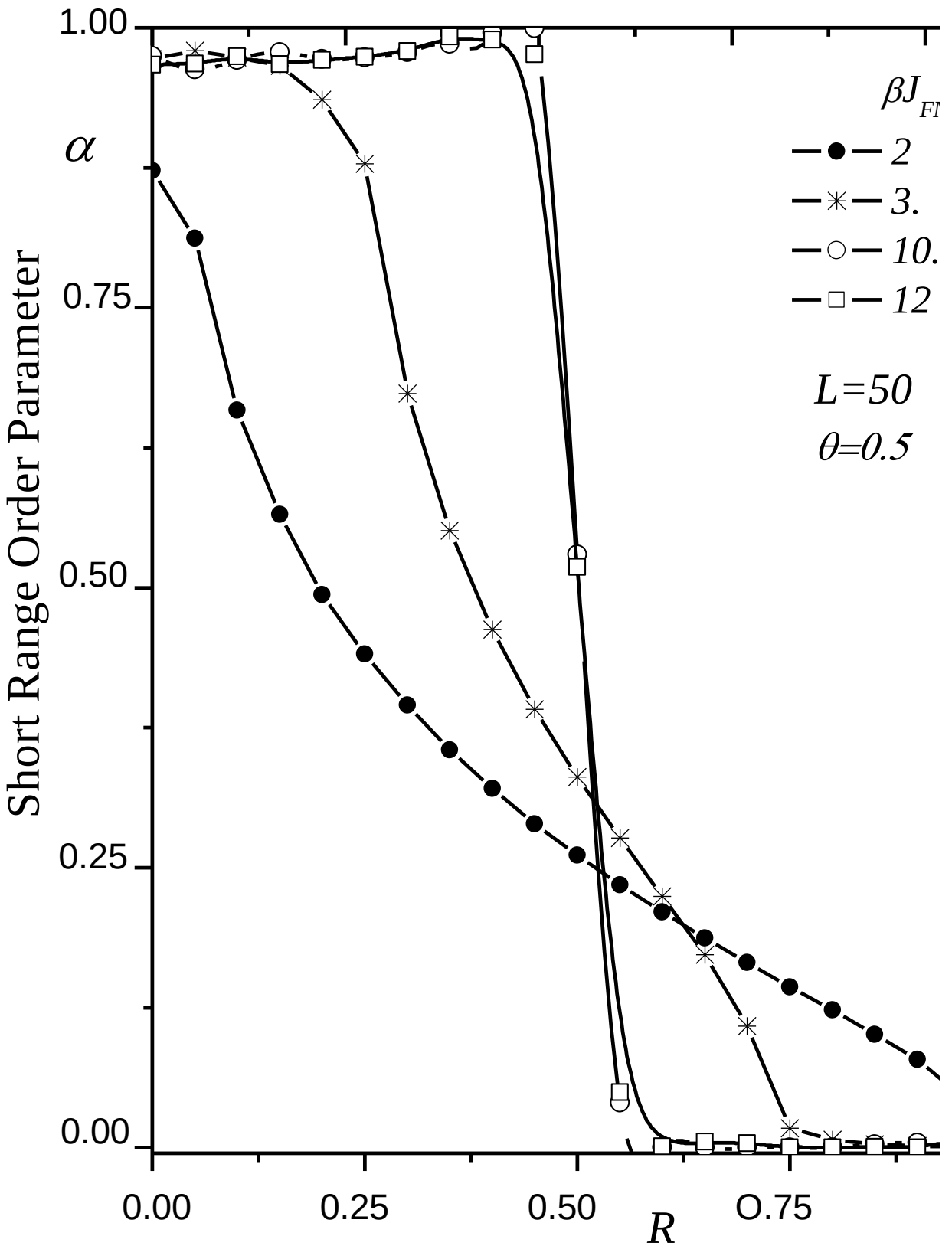

Fig. 4